\documentclass[onecolumn,conference]{IEEEtran}
\usepackage[utf8]{inputenc}
\usepackage{graphicx}
\usepackage{url}
\usepackage{subfig}
\usepackage{listings}
\begin{document}

\title{An Analysis of Publication Venues for\\Automatic Differentiation Research}
\author{\IEEEauthorblockN{Atılım Güneş Baydin and Barak A. Pearlmutter}\IEEEauthorblockA{Hamilton Institute \& Department of Computer Science\\National University of Ireland Maynooth, Maynooth, Co. Kildare, Ireland\\Email: atilimgunes.baydin@nuim.ie; barak@cs.nuim.ie}\\22 September 2014}

\maketitle

%\section*{Introduction}
We present the results of our analysis of publication venues for papers on automatic differentiation (AD), covering academic journals and conference proceedings.

Our data are collected from the \emph{AD publications database} maintained by the autodiff.org community website (\url{http://www.autodiff.org/}). The database is purpose-built for the AD field and is expanding via submissions by AD researchers. Therefore, it provides a relatively noise-free list of publications relating to the field. However, it does include noise in the form of variant spellings of journal and conference names. We handle this by manually correcting and merging these variants under the official names of corresponding venues. We share the raw data we get after these corrections, at the end of this report.

In the final data, we have 535 journal papers published between 1970--2014 and 410 conference papers published between 1979--2012.

\begin{figure}[h]
  \centering
  \subfloat[]{\includegraphics[width=8cm]{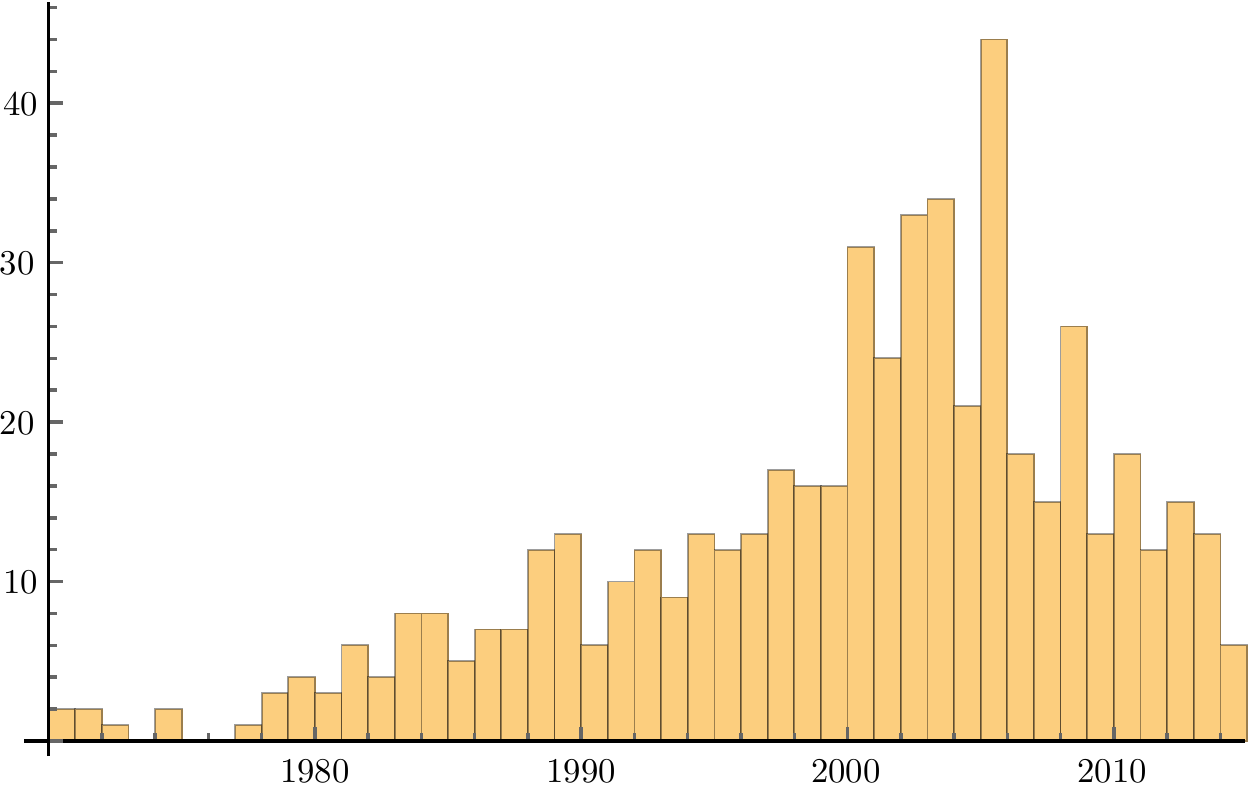}}\\
  %\qquad
  \subfloat[]{\includegraphics[width=8cm]{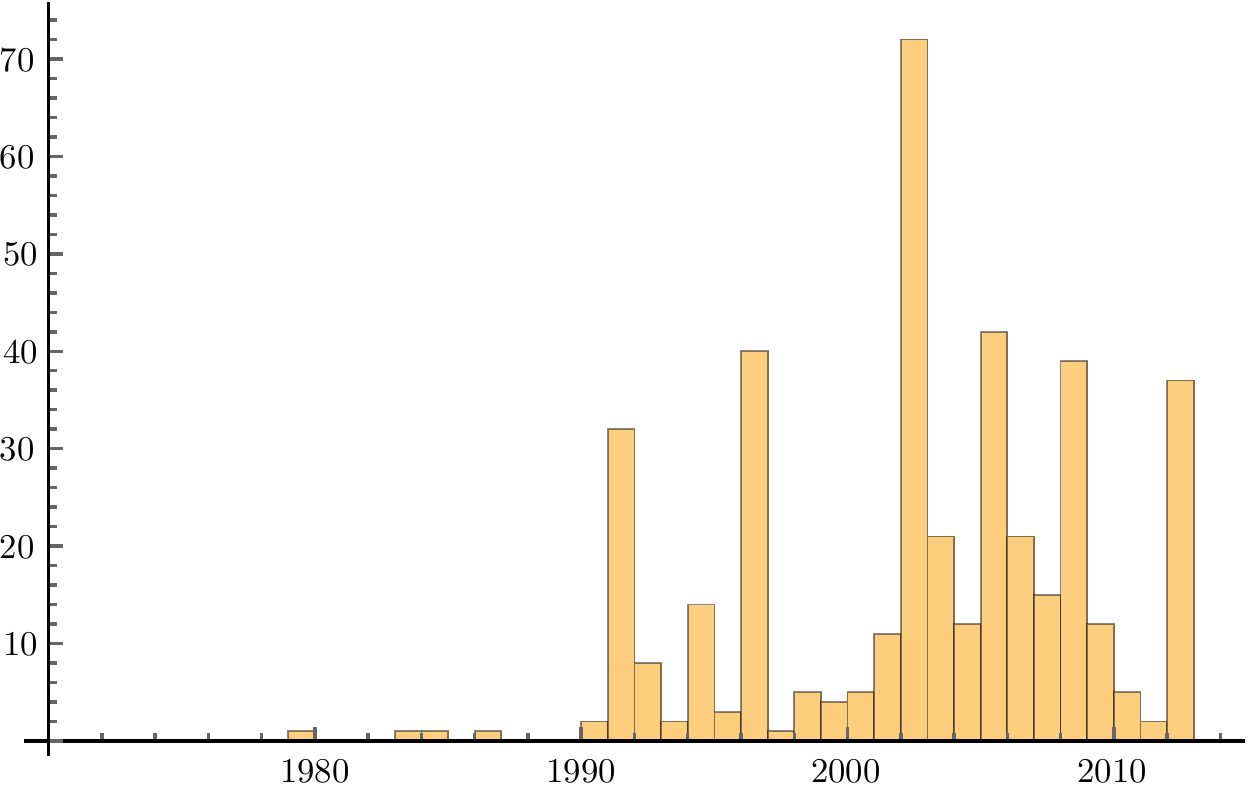}}
  \caption{Distribution of publication years of AD-related papers. (a) Journals; (b) Conference proceedings.}
  \label{FigureYears}
\end{figure}

\begin{figure*}
  \centering
  \includegraphics[width=\textwidth]{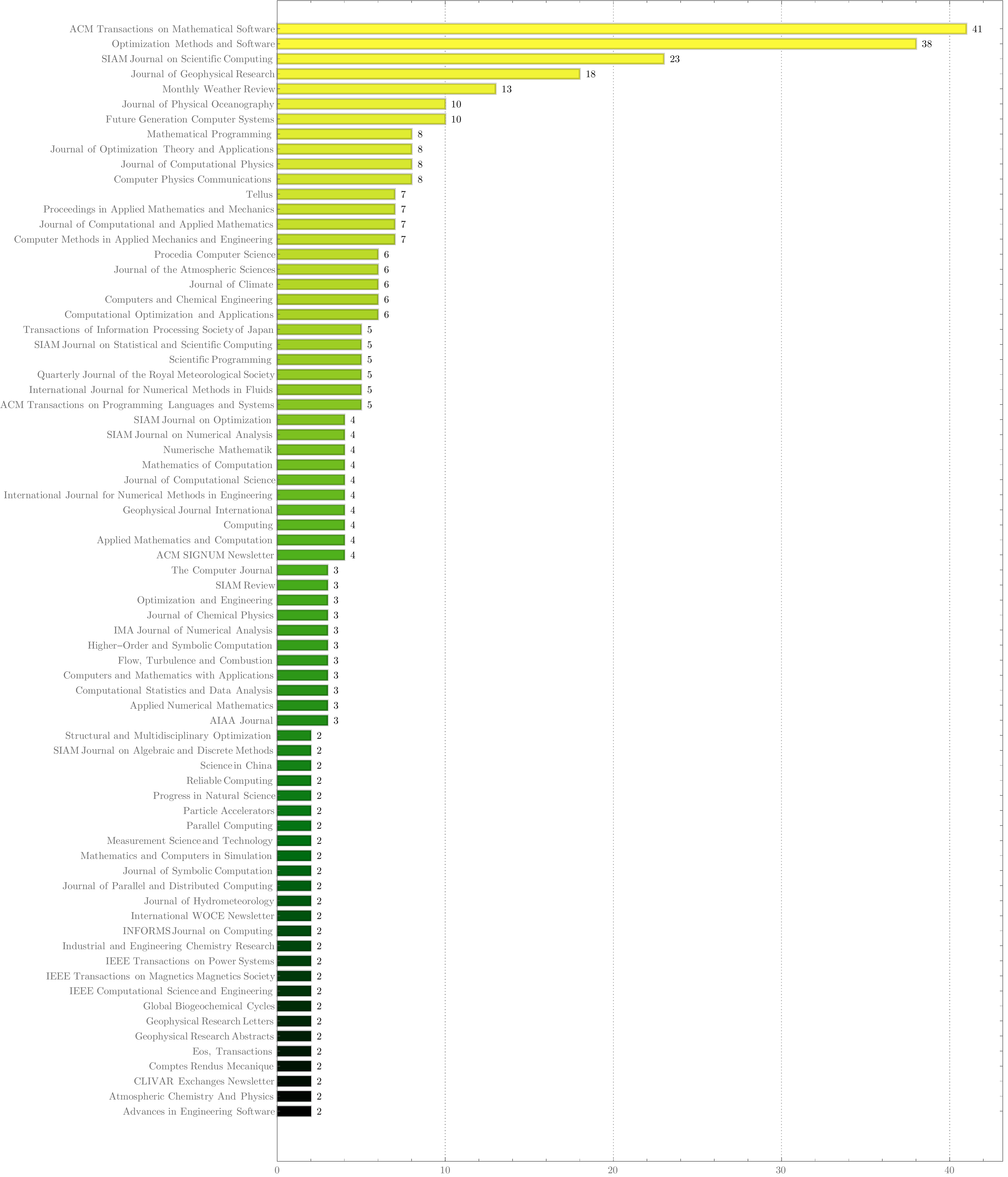}
  \caption{Ranking of journals by the number of AD-related published papers. Journals with less than 2 papers are not shown.}
  \label{FigureJournalsRanking}
\end{figure*}

\begin{figure*}
\parbox[c][20cm][c]{\textwidth}{
  \centering
  \includegraphics[width=\textwidth]{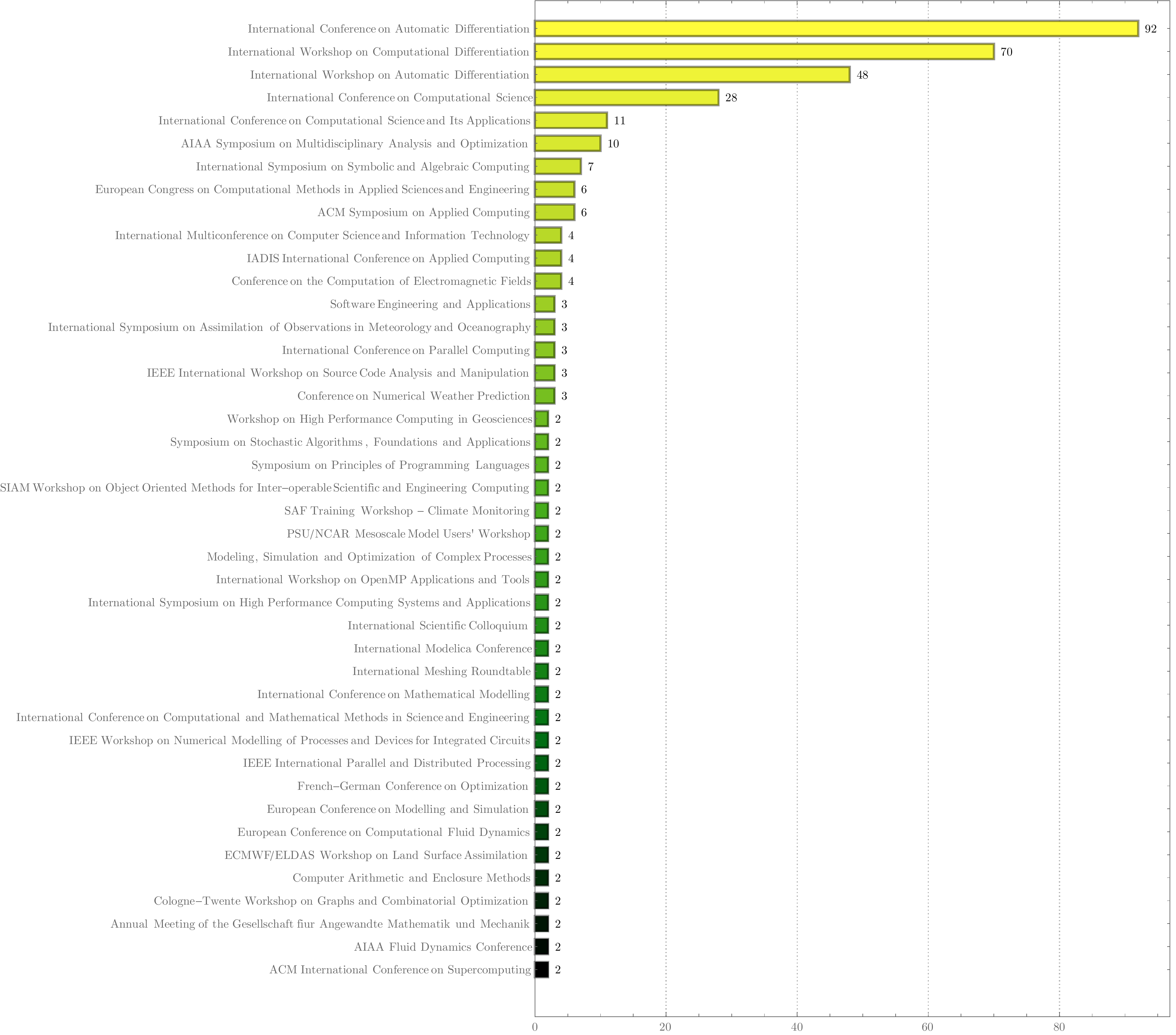}
  \caption{Ranking of conferences by the number of AD-related published papers. Proceedings with less than 2 papers are not shown.}
  \label{FigureConferencesRanking}
}
\end{figure*}

Figure~\ref{FigureYears} shows the distribution of publication years of AD-related papers. Figures \ref{FigureJournalsRanking} and \ref{FigureConferencesRanking} give the ranking of journals and conferences by the number of AD-related published papers. For keeping the size of these ranking figures manageable, we only include venues with 2 or more publications.

\newpage
\noindent Cleaned up raw data for journal papers:

\begin{lstlisting}
ACM Computing Surveys 1994, 1
ACM SIGNUM Newsletter 1987, 1; 1988, 1; 1989, 1; 1995, 1
ACM Transactions on Mathematical Software 1980, 2; 1981, 2; 1982, 2; 1984, 2; 1985, 1; 1992, 1; 1995, 2; 1996, 1; 1998, 1; 1999, 1; 2000, 5; 2002, 1; 2003, 2; 2004, 1; 2005, 2; 2006, 1; 2007, 1; 2008, 4; 2010, 1; 2012, 2; 2013, 4; 2014, 2
ACM Transactions on Programming Languages and Systems 1983, 1; 1989, 1; 1991, 1; 1993, 1; 2008, 1
Acta Crystallographica Section D  2004, 1
Acta Informatica  1972, 1
Acta Oecologica 2004, 1
Advances in Engineering Software  2000, 1; 2003, 1
Advances in Space Research  2000, 1
Advances in Water Resources 2006, 1
Aerosol Science And Technology  2005, 1
AIAA Journal  1997, 1; 2005, 1; 2008, 1
Annales Geophysicae 1996, 1
Annals of Biomedical Engineering  2010, 1
Applied Mathematics and Computation 1987, 2; 1990, 1; 2009, 1
Applied Numerical Mathematics 1992, 1; 1996, 1; 1997, 1
Atmospheric Chemistry And Physics 2003, 1; 2005, 1
Atmospheric Environment 1997, 1
Atmospheric Measurement Techniques  2011, 1
Automatisierungstechnik 2000, 1
Biometrics  1983, 1
BIT 2005, 1
Bulletin of the American Meteorological Society 1997, 1
BYTE  1986, 1
Canadian Journal of Fisheries and Aquatic Sciences  2001, 1
Chemical Engineering Science  2008, 1
Climatic Change 1997, 1
CLIVAR Exchanges Newsletter 2000, 2
CNLS Research Highlights  1998, 1
College Mathematics Journal 1989, 1
Communications in Nonlinear Science and Numerical Simulation  2013, 1
Complex Variables and Elliptic Equations  2009, 1
Comptes Rendus Mecanique  2008, 2
Computational Complexity  1997, 1
Computational Mathematics and Mathematical Physics  1997, 1
Computational Optimization and Applications 1995, 2; 1997, 1; 1998, 1; 2003, 1; 2004, 1
Computational Statistics and Data Analysis  1988, 1; 2006, 1; 2011, 1
Computer Graphics Forum 2013, 1
Computer Journal  1983, 1
Computer Methods in Applied Mechanics and Engineering 1994, 1; 1998, 1; 2005, 1; 2008, 1; 2009, 1; 2010, 1; 2011, 1
Computer Physics Communications 2000, 1; 2002, 1; 2005, 1; 2006, 1; 2007, 1; 2010, 2; 2013, 1
Computer Science and Information Systems  2007, 1
Computers and Chemical Engineering  1988, 1; 1990, 1; 1995, 1; 1998, 1; 2009, 1; 2013, 1
Computers and Fluids  2006, 1
Computers and Geosciences 2002, 1
Computers and Mathematics with Applications 1985, 1; 1986, 1; 1994, 1
Computers and Structures  2010, 1
Computers in Physics  1996, 1
Computing and Visualization in Science  2001, 1
Computing Systems in Engineering  1992, 1
Computing 1979, 1; 1986, 1; 1996, 1; 2012, 1
Current Science 2000, 1
Deep Sea Research Part I: Oceanographic Research Papers 2004, 1
Deep Sea Research Part II: Topical Studies in Oceanography  2011, 1
Discrete and Continous Dynamical Systems Supplements  2011, 1
Discrete Applied Mathematics  2008, 1
Discussiones Mathematicae, Differential Inclusions, Control and Optimization  2007, 1
Dynamical Systems and Differential Equations, AIMS  2013, 1
Earth and Planetary Science Letters 2007, 1
Econometrics Journal  2003, 1
Electrical Engineering in Japan 2005, 1
Electronic Notes in Theoretical Computer Science  2007, 1
Electronic Transactions on Numerical Analysis 2005, 1
Environmental and Ecological Statistics 2008, 1
Environmental Modelling and Software  2000, 1
Eos, Transactions 2002, 2
Finite Elements in Analysis and Design  1993, 1
Flow, Turbulence and Combustion 2000, 2; 2001, 1
Future Generation Computer Systems  2005, 10
GACM Report 2011, 1
General Relativity and Gravitation  2005, 1
Geophysical Journal International 2005, 1; 2006, 1; 2007, 2
Geophysical Research Abstracts  2003, 2
Geophysical Research Letters  2003, 1; 2006, 1
Geoscience and Remote Sensing Letters 2007, 1
Geoscientific Instrumentation, Methods and Data Systems 2013, 1
Global Biogeochemical Cycles  2002, 1; 2005, 1
Higher-Order and Symbolic Computation 2001, 1; 2008, 2
IEEE Computational Science and Engineering  1996, 1; 1998, 1
IEEE Transactions on Automatic Control  1994, 1
IEEE Transactions on Antennas and Propagation 2009, 1
IEEE Transactions on Electron Devices 1988, 1
IEEE Transactions on Geoscience and Remote Sensing  2008, 1
IEEE Transactions on Magnetics Magnetics Society  2010, 1; 2011, 1
IEEE Transactions on Power Systems  1997, 1; 1999, 1
IEICE Transactions  1991, 1
IMA Journal of Numerical Analysis 1986, 1; 1991, 1; 1992, 1
IMACS Annals of Computing and Applied Mathematics 1992, 1
IMACS Transactions on Scientific Computing  1989, 1
Industrial and Engineering Chemistry Research 1989, 1; 2004, 1
Informatik-Spektrum 2008, 1
INFORMS Journal on Computing  1997, 1; 2009, 1
International Journal for Numerical Methods in Engineering  1998, 1; 2003, 2; 2006, 1
International Journal for Numerical Methods in Fluids 2002, 1; 2005, 1; 2009, 1; 2010, 1; 2011, 1
International Journal of Computational Science and Engineering  2014, 1
International Journal of High Performance Computing and Networking  1999, 1
International Journal of High Speed Computing 2004, 1
International Journal of Numerical Methods in Fluids  2005, 1
International Journal of Radiation Oncology Biology Physics 1993, 1
International Journal on Control  1988, 1
International WOCE Newsletter 1999, 2
Japan Journal of Applied Mathematics  1984, 1
Journal of Applied Meteorology  2002, 1
Journal of Atmospheric And Oceanic Technology 2003, 1
Journal of Atmospheric and Solar-Terrestrial Physics  2004, 1
Journal of Atmospheric Chemistry  2005, 1
Journal of Chemical Physics 1992, 1; 2001, 1; 2005, 1
Journal of Climate  1997, 1; 2001, 1; 2002, 2; 2003, 2
Journal of Computational Acoustics  2004, 1
Journal of Computational and Applied Mathematics  1988, 1; 1991, 1; 1995, 1; 2000, 1; 2002, 1; 2008, 1; 2009, 1
Journal of Computational and Graphical Statistics 1995, 1
Journal of Computational Physics  1981, 1; 1994, 1; 1996, 2; 2000, 2; 2004, 1; 2005, 1; 2010, 1; 2013, 1; 2014, 2
Journal of Econometrics 1994, 1
Journal of Economic Dynamics and Control  1990, 1
Journal of Fluids Engineering 2007, 1
Journal of Geophysical Research 1998, 1; 1999, 3; 2000, 1; 2001, 3; 2002, 3; 2003, 2; 2004, 4; 2005, 1
Journal of Global Optimization  1995, 1
Journal of Graphics, GPU, and Game Tools  2004, 1
Journal of Hydrometeorology 2001, 1; 2003, 1
Journal of Information Processing 1991, 1
Journal of Marine Systems 2001, 1
Journal of Mathematical Analysis and Applications 1983, 1
Journal of Mathematical Physics 1981, 1
Journal of Object Oriented Programming  1990, 1
Journal of Oceanography 2006, 1
Journal of Optimization Theory and Applications 1979, 1; 1983, 1; 1984, 1; 1988, 1; 1989, 2; 2000, 1; 2004, 1
Journal of Parallel and Distributed Computing 1997, 1; 2008, 1
Journal of Physical Oceanography  2002, 4; 2003, 2; 2004, 1; 2005, 2; 2006, 1
Journal of Quantitative Spectroscopy and Radiative Transfer 2014, 1
Journal of Sound and Vibration  1995, 1
Journal of Symbolic Computation 1986, 1; 1994, 1
Journal of Systems and Software 2006, 1
Journal of the Atmospheric Sciences 1982, 1; 1996, 1; 1999, 1; 2003, 2; 2006, 1
Journal of the Institute of Mathematics and Applications  1974, 1
Journal of Volcanology and Geothermal Research  2008, 1
Journal of Water Resources Planning and Management  1999, 1
Journal on Numerical Methods and Computer Applications  2003, 1
Linear Algebra and Its Applications 1991, 1
Magnetics Transactions on 2005, 1
MATEKON 1985, 1
Matematicheskoe Modelirovanie 1989, 1
Mathematica Numerica Sinica 2009, 1
Mathematical and Computer Modelling 2003, 1
Mathematical Geology  2005, 1
Mathematical Modelling and Numerical Analysis 2002, 1
Mathematical Proceedings of the Cambridge Philosophical Society 1978, 1
Mathematical Programming  1978, 1984, 1; 1988, 1; 2002, 1; 2003, 1; 2004, 1; 2006, 1; 2008, 1
Mathematics and Computers in Simulation 2010, 1; 2011, 1
Mathematics Magazine  1986, 1
Mathematics of Computation  1984, 1; 1999, 1; 2000, 1; 2005, 1
Measurement Science and Technology  2002, 1; 2003, 1
Meteorology And Atmospheric Physics 2004, 1
Methoden Und Verfahren Der Mathematischen Physik  1971, 1
Monthly Weather Review  1992, 2; 1997, 1; 1999, 1; 2000, 1; 2001, 2; 2002, 4; 2003, 1; 2005, 1
Neural Computation  1994, 1
Nonlinear Optimization  1981, 1
Numerical Algorithms  1998, 1
Numerische Mathematik 1977, 1; 1982, 1; 1993, 1; 1994, 1
Ocean Modeling Journal  2006, 1
Optimization and Engineering  2001, 1; 2005, 1; 2008, 1
Optimization Methods and Software 1991, 1; 1992, 2; 1993, 2; 1994, 2; 1996, 1; 1997, 1; 1998, 4; 1999, 2; 2000, 2; 2001, 2; 2002, 2; 2003, 1; 2009, 1; 2010, 1; 2011, 1; 2012, 12; 2013, 1
Optimization  2001, 1
ORSA Journal on Computing 1993, 1
Parallel Computing  1990, 1; 2001, 1
Particle Accelerators 1989, 2
Physica D.  2005, 1
Physics and Chemistry of the Earth  1996, 1
Physics of Fluids 2005, 1
Policy Analysis and Information Systems 1979, 1
Procedia Computer Science 2010, 4; 2011, 1; 2013, 1
Proceedings in Applied Mathematics and Mechanics  2003, 3; 2007, 3; 2010, 1
Progress in Natural Science 2002, 1; 2009, 1
Quarterly Journal of the Royal Meteorological Society 1987, 2; 2001, 1; 2002, 1; 2003, 1
Reliable Computing  1998, 1; 2007, 1
Remote Sensing of the Environment 2006, 1
Science in China  2004, 1; 2009, 1
Science 2001, 1
Scientific Programming  1992, 1; 2001, 1; 2003, 1; 2006, 1; 2008, 1
SIAG/OPT Views-and-News 2000, 1
SIAM Journal of Numerical Analysis  1978, 1
SIAM Journal on Algebraic and Discrete Methods  1983, 1; 1986, 1
SIAM Journal on Computing 1988, 1
SIAM Journal on Discrete Mathematics  1981, 1
SIAM Journal on Matrix Analysis and Applications  1996, 1
SIAM Journal on Numerical Analysis  1974, 1; 1979, 1; 1983, 1; 1997, 1
SIAM Journal on Optimization  1991, 1; 2002, 2; 2005, 1
SIAM Journal on Scientific Computing  1994, 2; 1995, 1; 1997, 1; 1998, 2; 1999, 2; 2000, 3; 2001, 1; 2002, 1; 2003, 1; 2005, 3; 2006, 1; 2007, 1; 2008, 3; 2009, 1
SIAM Journal on Statistical and Scientific Computing  1980, 1; 1987, 1; 1988, 1; 1992, 1; 1993, 1
SIAM News 1991, 1
SIAM Review 1988, 1; 2005, 1; 2010, 1
SIGACT News 1985, 1
Simulation Modelling Practice and Theory  2004, 1
Software--Practice and Experience 1997, 1
Structural and Multidisciplinary Optimization 1995, 1; 2000, 1
Studia Geophysica et Geodaetica 2001, 1
Tellus  1985, 1; 1990, 1; 1993, 1; 1996, 1; 2000, 2; 2003, 1
The American Statistician 1984, 1
The Computer Journal  1970, 2; 2008, 1
The Journal of Chemical Physics 2001, 1
The Review of Economics and Statistics  1984, 1
Theoretical Computer Science  1983, 1
Transactions of Information Processing Society of Japan 1987, 1; 1988, 1; 1989, 3
Transactions of the American Nuclear Society  2010, 1
Transactions on Graphics  2011, 1
Weather and Forecasting 2002, 1
Wissenschaftliche Zeitschrift der Technischen Hochschule fuer Chemie, Leuna-Merseburg  1971, 1
WSEAS Transactions on Circuits and Systems  2005, 1
WSEAS Transactions on Mathematics 2006, 1
Yugoslav Journal of Operations Research 1991, 1
\end{lstlisting}

\newpage
\noindent Cleaned up raw data for conference papers:

\begin{lstlisting}
Aachen Symposium on Signal Theory 2001, 1
ACM International Conference on Supercomputing  2001, 2
ACM SIGPLAN International Conference on Functional Programming  1998, 1
ACM SIGPLAN Workshop on Partial Evaluation and Semantics-Based Program Manipulation 2002, 1
ACM Symposium on Applied Computing  2002, 1; 2003, 3; 2004, 1; 2009, 1
ACM Symposium on Parallel Algorithms and Architectures  1999, 1
ACM Symposium on the Theory of Computing  1983, 1
Advances in Reactor Physics 2012, 1
AIAA Aerospace Sciences Meeting and Exhibit 2001, 1
AIAA Computational Fluid Dynamics Conference  1993, 1
AIAA Fluid Dynamics Conference  1994, 2
AIAA Symposium on Multidisciplinary Analysis and Optimization 1992, 2; 1994, 7; 1998, 1; 2012, 1
Annual Meeting of the Gesellschaft fiur Angewandte Mathematik und Mechanik  2002, 2
Array Conference  1986, 1
Colloque Africain sur la Recherche en Informatique  2002, 1
Cologne-Twente Workshop on Graphs and Combinatorial Optimization  2009, 2
Computer Arithmetic and Enclosure Methods 1992, 2
Computers and Information in Engineering Conference 2007, 1
Conference on Atmospheric and Oceanic Fluid Dynamics  2003, 1
Conference on Engineering Applications and Computational Algorithms 2003, 1
Conference on Numerical Weather Prediction  1998, 1; 2001, 2
Conference on Scientific Computing  2000, 1
Conference on the Computation of Electromagnetic Fields 2007, 2; 2009, 2
Conference on Weather Analysis and Forecasting  2005, 1
Dagstuhl Seminar on Combinatorial Scientific Computing  2009, 1
ECMWF/ELDAS Workshop on Land Surface Assimilation 2005, 2
EUROMECH Nonlinear Dynamics Conference  2005, 1
European Conference on Computational Fluid Dynamics 2006, 2
European Conference on Modelling and Simulation 2006, 2
European Congress on Computational Methods in Applied Sciences and Engineering  1996, 1
European Congress on Computational Methods in Applied Sciences and Engineering  2004, 5
European Simulation Multiconference 2004, 1
EUROSIM Congress on Modelling and Simulation  2001, 1
French-German Conference on Optimization  1994, 2
High Performance Computing Symposium  1995, 1
IADIS International Conference on Applied Computing 2007, 4
IASTED International Conference on Modelling, Identification, and Control 2002, 1
IEEE Conference on Decision and Control 2012, 1
IEEE International Conference on Cluster Computing  2010, 1
IEEE International Conference on Computer Design  1992, 1
IEEE International Parallel and Distributed Processing  1998, 2
IEEE International Workshop on Parallel and Distributed Scientific and Engineering Computing  2009, 1
IEEE International Workshop on Source Code Analysis and Manipulation  2002, 1; 2004, 1; 2006, 1
IEEE Radar Conference 2010, 1
IEEE Workshop on Numerical Modelling of Processes and Devices for Integrated Circuits 1992, 2
International Carbon Dioxide Conference 2001, 1
International Conference of Modeling and Simulation 2008, 1
International Conference on Architecture of Computing Systems 2002, 1
International Conference on Automatic Differentiation 2005, 30; 2008, 31; 2012, 31
International Conference on Computational and Mathematical Methods in Science and Engineering 2010, 2
International Conference on Computational Science 2001, 1; 2002, 10; 2003, 2; 2004, 1; 2005, 1; 2006, 11; 2008, 1
International Conference on Computational Science and Its Applications  2003, 8; 2006, 2; 2009, 1
International Conference on Fuzzy Systems and Knowledge Discovery 2009, 1
International Conference on Heat and Mass Transfer  2009, 1
International Conference on Mathematical Modelling  2003, 1; 2012, 1
International Conference on Metal Forming 2012, 1
International Conference on Parallel and Distributed Processing Techniques and Applications 2001, 1
International Conference on Parallel Computing  2007, 1; 2008, 2
International Conference on Solid-State Sensors, Actuators and Microsystems 2005, 1
International Conference on Underwater System Technology  2008, 1
International Congress on Mathematical Software 2006, 1
International Euro-Par Conference 2000, 1
International Federation for Information Processing Conference  1984, 1
International Meshing Roundtable  2002, 1; 2004, 1
International Modelica Conference 2008, 1; 2012, 1
International MultiConference of Engineers and Computer Scientists  2009, 1
International Multiconference on Computer Science and Information Technology  2007, 3; 2009, 1
International Scientific Colloquium 2000, 1; 2011, 1
International Symposium on Assimilation of Observations in Meteorology and Oceanography 1990, 2; 1995, 1
International Symposium on High Performance Computing Systems and Applications  2002, 1; 2003, 1
International Symposium on Scientific Computing, Computer Arithmetic and Validated Numerics 1996, 1
International Symposium on Symbolic and Algebraic Computing 1992, 1; 1993, 1; 1997, 1; 1999, 1; 2002, 1; 2003, 1; 2007, 1
International Symposium on Symbolic and Numeric Algorithms for Scientific Computing 2004, 1
International Workshop on Aircraft Systems Technologies 2007, 1
International Workshop on Automatic Differentiation 2002, 48
International Workshop on Computational Differentiation 1991, 32; 1996, 38
International Workshop on Equation-Based Object-Oriented Modeling Languages and Tools 2011, 1
International Workshop on Implementation and Application of Functional Languages  2005, 1
International Workshop on OpenMP Applications and Tools 2005, 2
Macsyma User's Conference 1979, 1
Modeling, Simulation and Optimization of Complex Processes  2005, 1; 2008, 1
Modelling, Identification, and Control  2003, 1
North American Power Symposium  1994, 1
PSU/NCAR Mesoscale Model Users' Workshop  2002, 2
Python for Scientific Computing Conference  2010, 1
SAF Training Workshop - Climate Monitoring  2000, 2
Scientific Discovery through Advanced Computing Program 2005, 1
SIAM Workshop on Object Oriented Methods for Inter-operable Scientific and Engineering Computing  1999, 2
Software Engineering and Applications 2002, 1; 2003, 1; 2004, 1
Symposium on Principles of Programming Languages  2007, 2
Symposium on Simulation Techniques  2005, 1
Symposium on Stochastic Algorithms, Foundations and Applications  2001, 1; 2003, 1
Workshop on Applied Parallel Computing in Physics,  Chemistry and Engineering Science 1996, 1
Workshop on Automatic Data Layout and Performance Prediction  1995, 1
Workshop on High Performance Computing in Geosciences 1994, 2
Workshop on Parallel Systems and Algorithms 2008, 1
Workshop on Symbolic and Numeric Computing  1994, 1
World Congress on Intelligent Control and Automation  2006, 1
\end{lstlisting}

\end{document}